# Direct observation of the Dzyaloshinskii-Moriya interaction in a Pt/Co/Ni film

Kai Di, Vanessa Li Zhang, Hock Siah Lim, Ser Choon Ng, Meng Hau Kuok[a]

*Department of Physics, National University of Singapore, Singapore 117551*

Jiawei Yu, Jungbum Yoon, Xuepeng Qiu, and Hyunsoo Yang[b]

*Department of Electrical and Computer Engineering, National University of Singapore,*

*Singapore 117576*

The interfacial Dzyaloshinskii-Moriya interaction (DMI) in an in-plane anisotropic Pt(4nm)/Co(1.6nm)/Ni(1.6nm) film has been directly observed by Brillouin spectroscopy. It is manifested as the asymmetry of the measured magnon dispersion relation, from which the DMI constant has been evaluated. Linewidth measurements reveal that the lifetime of the magnons is asymmetric with respect to their counter-propagating directions. The lifetime asymmetry is dependent on the magnon frequency, being more pronounced the higher the frequency. Analytical calculations of the magnon dispersion relation and linewidth agree well with experiments.

PACS numbers: 75.30.Ds, 75.70.Ak, 75.70.Tj, 78.35.+c

[a] phykmh@nus.edu.sg.

[b] eleyang@nus.edu.sg.

The Dzyaloshinskii-Moriya interaction (DMI), an anti-symmetric exchange interaction, was initially proposed to explain the weak ferromagnetism of antiferromagnets both phenomenologically [1] and microscopically [2]. Recently, DMI has attracted renewed interest due to its fundamental role in various novel phenomena such as the magnon Hall effect [3], molecular magnetism [4] and multiferroicity [5]. Also, DMI is now known to be responsible for chiral spin textures such as magnetic skyrmions [6-8], which show promise in spintronics and high-speed, ultra-high density storage technology due to their unique properties like propagation under ultralow current densities [9, 10] and rewritability by spin-polarized currents [11].

DMI can arise from inversion symmetry breaking at surfaces or interfaces between a ferromagnetic layer and a nonmagnetic one having a strong spin-orbit coupling [12-15] due to 3-site indirect exchange mechanism [16, 17]. Such an interfacial DMI may exist regardless of the crystal symmetry of the component materials [18], and is expected to be much stronger than the bulk interaction [12, 13]. Among these layered structures, Pt/Co is of enormous scientific and technological interest because of its strong perpendicular magnetic anisotropy (PMA), which results in high thermal stability and reduced threshold current for spin-transfer switching [19] and current-driven domain wall (DW) motion [20, 21]. The interfacial DMI can be tuned, by interface engineering, to change the type (Néel or Bloch), chirality, and velocity of the DWs [22-24]. Because of the significance of Pt/ferromagnet contacts in spintronics [25], it is critical to study the spin dynamics of such interfaces in the presence of DMI, especially when spin waves (SWs) serve as spin-current carriers. However, no direct experimental evidence of the interfacial DMI in Pt/ferromagnet films has been reported.

While the momentum-resolved spin dynamics measurement techniques of spin-polarized electron energy loss spectroscopy [15] and inelastic neutron scattering [26] can offer direct evidence of the presence of DMI, they are not suitable for multilayers with buried

Pt/ferromagnet thin films [18]. Magnons probed by these methods generally have energies, lifetimes and attenuation lengths of the order of 10 meV, 10 femtoseconds, and 1 nm [27], respectively. However, for magnonics and spintronics applications, SWs in the GHz range (0.01 to 0.1 meV), which can be readily excited by microwaves or spin transfer torque [28, 29], are of greater relevance due to their longer lifetime in the nanosecond range and micron-scale coherence length. Within this energy scale, Brillouin light scattering (BLS) with its sub-GHz resolution and high surface sensitivity (~ few nm) for metals is ideally suited for studying both the interfacial DMI and spin-wave dynamics of Pt/ferromagnet multilayers.

We report on a BLS study of interfacial DMI in an as-grown Pt/Co/Ni multilayer film possessing an in-plane magnetization. DMI-induced effects such as the asymmetry of the magnon dispersion relations and spectral linewidths were observed. Experiments also reveal that the asymmetry of magnon lifetime is dependent on its frequency. Analytical macroscopic calculations of the spin-wave dispersion relation and linewidth are in agreement with the measured data.

The sample studied was deposited on a thermally oxidized silicon wafer by both DC and RF magnetron sputtering at room temperature. Specifically, the unannealed film stack is substrate/MgO(2)/Pt(4)/Co(1.6)/Ni(1.6)/MgO(2)/SiO$_2$(3) (hereafter referred to as sample Pt/Co/Ni), where the figures in parentheses are the nominal thicknesses in nm. Argon gas (~2.3 mTorr) was used during the sputtering process with a background pressure of $2\times10^{-9}$ Torr, and the deposition rates for MgO, Pt, Co, Ni and SiO$_2$ were 0.026, 0.54, 0.14, 0.21 and 0.10 Å/s, respectively [30]. The in-plane and out-of-plane magnetic hysteresis loops of the sample, measured by vibrating sample magnetometry (VSM), as shown in Fig. 1, reveal that the in-plane saturation field and the saturation magnetization $M_S$ are approximately 50 mT/$\mu_0$ and 1160 kA/m, respectively.

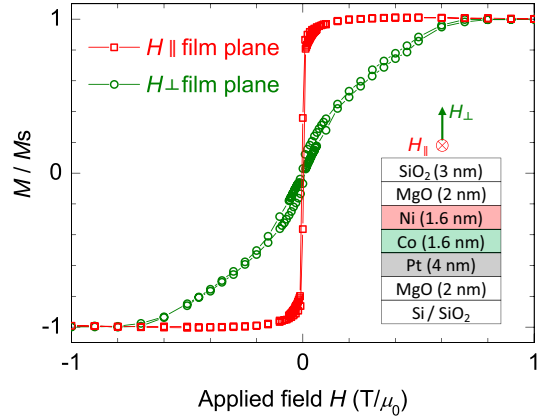

FIG. 1. In-plane and perpendicular magnetic hysteresis loops of the Pt(4)/Co(1.6)/Ni(1.6) film measured by VSM. Inset: Schematic of the multilayer film and directions of the applied magnetic fields.

All BLS measurements were performed in the 180° back-scattering geometry (see inset of Fig. 2) and in *ps* polarization using the 514.5nm radiation of an argon-ion laser and a six-pass tandem Fabry-Perot interferometer. The in-plane DC magnetic field $H_0$ was applied perpendicular to the incident plane of light, corresponding to the Damon-Eschbach (DE) geometry. As the metallic film is very thin, counter-propagating surface waves localized at the top and bottom interfaces were simultaneously observed in the BLS experiments. In the light scattering process, as total momentum is conserved along the film surface, the Stokes (magnon creation) and anti-Stokes (magnon annihilation) peaks arise from SWs propagating in the –*x* and +*x* directions, respectively. Thus, the respective frequencies of counter-propagating SWs, having the same momentum, are simultaneously presented in the same spectrum. Figure 2 shows that the magnon peaks in the Stokes and anti-Stokes portions of a typical spectrum are asymmetric in terms of both frequency and intensity. It is noteworthy that on reversing the direction of the applied magnetic field $H_0$, the respective center frequencies, linewidths and intensities of the Stokes and anti-Stokes peaks were also interchanged. This is a consequence of the reversed spin-wave propagation direction, as reversing the magnetization is equivalent to a time-reversal operation [27].

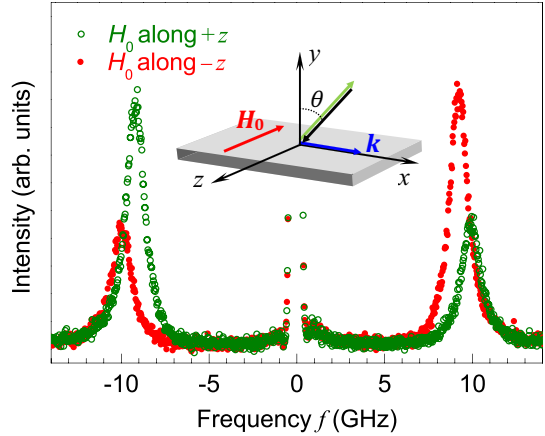

FIG. 2. Brillouin spectra recorded at a fixed incident angle of $\theta = 45°$ ($|k|=17.3$ μm$^{-1}$) under oppositely-oriented external magnetic fields $H_0 = 97\,\text{mT}/\mu_0$. Inset: Schematic of the Cartesian coordinate system and $180°$ back-scattering geometry. The incident and scattered light beams lie in the *x-y* plane and are at an angle $\theta$ to the *y*-axis. The magnon wave vector is denoted by $k$.

In the following, the BLS measurements were performed with the magnetization saturated along the –*z* direction under magnetic fields $H_0 > 50\,\text{mT}/\mu_0$. The momentum transfer in the film plane was varied by changing the incident angle of the laser beam, thus allowing the recording of BLS spectra at different wavevectors [31]. The resulting spin-wave dispersion relations measured under various applied magnetic fields are presented in Fig. 3(a). The most prominent feature of the dispersion curves is their asymmetry with respect to the wave vector $k$, which is attributed to the lifted chiral degeneracy arising from the interfacial DMI [14]. For the same wavelength, the frequency and group velocity of SWs propagating in the –*x* direction are larger than those of SWs propagating in the +*x* direction.

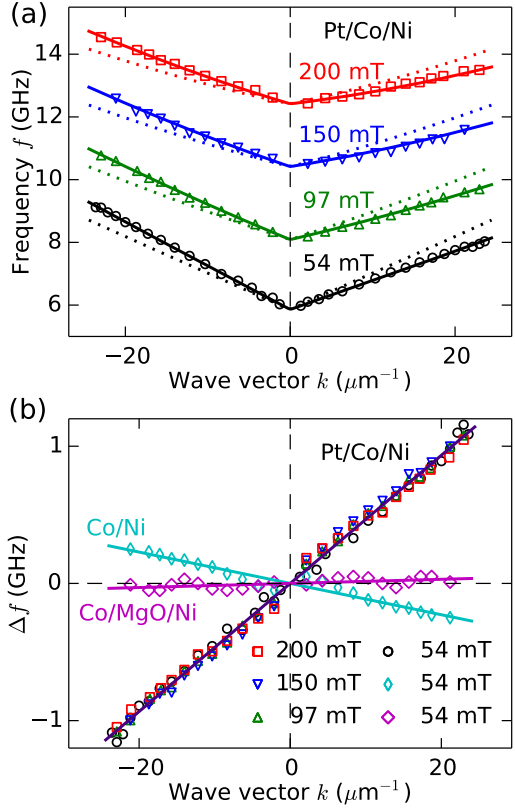

FIG. 3. (a) Spin-wave dispersion relations of the Pt/Co/Ni film at various applied magnetic fields. (b) Frequency difference of counter-propagating SWs as a function of wave vector. Symbols denote measured data. The symmetrical dotted lines in (a) represent the theoretical dispersion curves in the absence of DMI, while solid lines in (a) and (b) represent respective fits to Eqs. (6) and (7).

In the following quantitative analysis, the magnetic film was considered to be an effective uniform medium [32, 33], since the Co and Ni layers are coupled by strong exchange interactions and only long-wavelength acoustic SWs are of interest. Phenomenologically, the DMI energy density of our sample is given by [13, 34, 35]

$$E_{\text{DM}} = \frac{D}{M_S^2}\left(M_y \frac{\partial M_x}{\partial x} - M_x \frac{\partial M_y}{\partial x} + M_y \frac{\partial M_z}{\partial z} - M_z \frac{\partial M_y}{\partial z}\right), \quad (1)$$

where $D$ is the DMI constant, and $M_i$ the $i$-component of the magnetization. The magnetization of linear SWs can be expressed as $\boldsymbol{M}(x,t) = M_S \boldsymbol{m}(x,t)$. The unit vector along the magnetization direction is

$$\boldsymbol{m}(x,t) = m_x \hat{\boldsymbol{e}}_x + m_y \hat{\boldsymbol{e}}_y - \hat{\boldsymbol{e}}_z = m_{x0} e^{i(\omega t - kx)} \hat{\boldsymbol{e}}_x + m_{y0} e^{i(\omega t - kx)} \hat{\boldsymbol{e}}_y - \hat{\boldsymbol{e}}_z, \tag{2}$$

where $|m_{x0}|, |m_{y0}| \ll 1$, and $\hat{\boldsymbol{e}}_x, \hat{\boldsymbol{e}}_y, \hat{\boldsymbol{e}}_z$ are unit vectors codirectional with the coordinate axes. For SWs traveling along the $x$-axis, the effective field due to the DMI is given by [35, 36]

$$\boldsymbol{H}_{DM} = -\frac{1}{\mu_0}\frac{\delta E_{DM}}{\delta \boldsymbol{M}} = \frac{2D}{\mu_0 M_S}\left(\frac{\partial m_y}{\partial x}\hat{\boldsymbol{e}}_x - \frac{\partial m_x}{\partial x}\hat{\boldsymbol{e}}_y\right). \tag{3}$$

The total effective field is

$$\boldsymbol{H}_{eff} = -H_0 \hat{\boldsymbol{e}}_z + J\nabla^2 \boldsymbol{m} + \boldsymbol{H}_{dip} + \boldsymbol{H}_{ani} + \boldsymbol{H}_{DM}, \tag{4}$$

where $J = 2A/(\mu_0 M_S)$, $A$ is the exchange stiffness constant, the dipolar field in the film [37] is $\boldsymbol{H}_{dip} = -M_S \xi(kL) m_x \hat{\boldsymbol{e}}_x - M_S[1 - \xi(kL)] m_y \hat{\boldsymbol{e}}_y$, $L$ (= 3.2 nm) is the magnetic film thickness, $\xi(x) = 1 - (1 - e^{-|x|})/|x|$, the perpendicular uniaxial anisotropic field $\boldsymbol{H}_{ani} = 2K/(\mu_0 M_S) m_y \hat{\boldsymbol{e}}_y$, and $K$ is the anisotropy constant. Substituting Eqs. (2) and (4) into the Landau-Lifshitz-Gilbert (LLG) equation

$$\frac{d\boldsymbol{m}}{dt} = -\mu_0 \gamma \boldsymbol{m} \times \boldsymbol{H}_{eff} + \alpha \boldsymbol{m} \times \frac{d\boldsymbol{m}}{dt} \tag{5}$$

yields the following spin-wave dispersion relation

$$\omega = \omega_0 + \omega_{DM} = \mu_0 \gamma \sqrt{\left[H_0 + Jk^2 + \xi(kL)M_S\right]\left[H_0 - H_U + Jk^2 + M_S - \xi(kL)M_S\right]} - \frac{2\gamma}{M_S}Dk, \tag{6}$$

where the Gilbert damping constant $\alpha$ has been set to zero, $\gamma$ is the gyromagnetic ratio, and $H_U = 2K/(\mu_0 M_S)$. The angular frequency in the absence of DMI is denoted by $\omega_0$, while the DMI-induced frequency shift by $\omega_{DM}$. Fitting the experimental data to Eq. (6), based on the VSM-measured value of $M_S$, yielded the following values of the magnetic parameters: $\gamma = 194 \pm 2$ GHz/T (corresponding to a g-factor of $\approx$ 2.2), $H_U = 670 \pm 20$ kA/m, $D = 0.44 \pm 0.02$ mJ/m$^2$, and $A = 17 \pm 1$ pJ/m. Assuming a second-order perpendicular uniaxial anisotropy, the out-of-plane saturation field was estimated to be $H_{SAT,\perp} = M_S - H_U = 0.61$ T/$\mu_0$, which is of the order of the VSM measured value of about 0.7 T/$\mu_0$. Consideration of the anisotropic field in the calculations is necessary for ultrathin Pt/Co structures, which usually possess perpendicular magnetic anisotropy. The DMI-induced frequency difference of counter-propagating SWs, given by

$$\Delta f(k) = [\omega(-k) - \omega(k)]/2\pi = \frac{2\gamma}{\pi M_S} Dk, \quad (7)$$

is linear in the DMI constant and the spin-wave wave vector, but independent of the applied in-plane magnetic field. Figure 3(b) reveals that this relationship is borne out by the experimental Brillouin data. It is noteworthy that the frequency difference is more pronounced the shorter the wavelength, which is the case for exchange SWs [15]. Such a feature would be desirable in the nano-miniaturization of integrated magnonic circuits based on Pt/Co structures.

We now discuss why the observed asymmetry is not due to the nonreciprocity of surface SWs. For the scattering geometry shown in Fig. 2, the dynamic magnetizations of +$k$ SWs tend to localize at the top surface, while those of –$k$ SWs at the bottom surface [38], which is consistent with the higher BLS intensities of the former. Consequently, asymmetric distributions of magnetic parameters across the sample's thickness could lead to a frequency shift for counter-propagating SWs. First, the PMA is asymmetric as it mainly originates from the bottom Pt/Co interface. Due to the nonreciprocal localization of SWs, the +$k$ SW generally experiences an average anisotropy field $H_U$ weaker than that felt by the –$k$ SW. However, according to Eq. (6), the frequency of the former should be higher than that of the latter, which contradicts our observation. Thus, asymmetric PMA cannot explain the observed asymmetry. Second, the saturation magnetization $M_S$ is asymmetric due to the Co/Ni bilayer structure. To study its effect, two more samples were fabricated under the same conditions as for the Pt/Co/Ni sample. They are substrate/MgO(2)/Co(1.6)/Ni(1.6)/MgO(2)/SiO2(3) and substrate/MgO(2)/Co(1.6)/MgO(0.4)/Ni(1.6)/MgO(2)/SiO2(3), referred to as samples Co/Ni and Co/MgO/Ni respectively. The measured $\Delta f$ of the Co/MgO/Ni sample [see Fig. 3(b)] is very small, indicating that the contribution of the asymmetric distribution of $M_S$ is insignificant. This also justifies the use of the above effective medium theory. We have also to consider the asymmetry of $M_S$ arising from the proximity-induced magnetic moment in the Pt layer of the Pt/Co/Ni sample. However, because the magnetization of Pt is highly localized within the first few atomic layers adjacent to the Pt/Co interface and is much weaker than those of Co and Ni [39], the resulting frequency asymmetry should be much weaker than that arising from the asymmetric $M_S$ due to the Co/Ni structure. Therefore, such a proximity effect cannot account for the observed asymmetry of the Pt/Co/Ni sample. Interestingly, the measured $\Delta f$ of the Co/Ni sample [see Fig. 3(b)] is fairly large, suggesting the existence of asymmetric surface anisotropy and/or DMI at the Co/Ni interface, similar to the case of Fe/Ni [40]. However, since the $\Delta f$ of the Co/Ni sample has a sign opposite to that of the Pt/Co/Ni sample, the factors contributing to dispersion asymmetry in the former sample are not the same as those for the latter. Hence, we conclude that the observed strong asymmetric dispersion curve of the Pt/Co/Ni sample is principally due to the existence of DMI at the Pt/Co interface.

The effect of interfacial DMI on the magnon lifetime was also investigated by analyzing the spin-wave linewidth, namely the full width at half maximum (FWHM) of the Brillouin peak. For the linewidth study, additional spectra were recorded at various wave vectors in the DE geometry under $H_0 = 54$ mT/$\mu_0$. Three factors contribute to the measured linewidth: the finite lifetime of magnons due to damping, the instrumental broadening due to the limited resolving power of the optics, and the finite size of the collection aperture [41]. To obtain the true spin-wave linewidth (lineshape assumed to be Lorentzian), both instrumental broadening and aperture effects were considered in the spectral fitting, which was carried out following the procedure of Ref. 41. The instrumental function of our optical system used was a Gaussian function with a FWHM of ≈ 0.17 GHz, which was obtained by measuring the elastic Rayleigh peak and assuming the laser line, with a linewidth of a few MHz, to be a delta function. The aperture effect was reduced by using an f/7 lens, which served as the focusing and collection lens. Each spectrum was recorded for a Brillouin peak intensity of at least one thousand counts, which typically entailed a scanning duration of 10 hours for a laser-light power of 90 mW incident on the sample.

The deconvoluted linewidths of the counter-propagating SWs, displayed in Fig. 4, indicate that their lifetimes are asymmetric with respect to their propagation direction. This is attributed to the DMI at the Pt/Co interface. From Fig. 4(a), one observes that for the same wavelength, the linewidth of SWs propagating in the –x direction is wider than that of SWs in the +x direction. To eliminate the contribution of the frequency difference to the linewidth asymmetry, the same set of measured linewidth data are re-presented in Fig. 4(b) by plotting them against frequency. We see that for a given frequency, the +k SW generally has a narrower linewidth and hence longer lifetime than those of the –k SW. Also, the linewidth difference between counter-propagating SWs increases with increasing frequency. For $H_0 = 54$ mT/$\mu_0$, when the wave vector approaches zero and the frequency reaches a minimum of about 6 GHz [see Fig. 3(a)], the linewidth difference vanishes. Of note is that, in the scattering configuration of Fig. 2, when the direction of the applied magnetic field was reversed, the linewidths of the –k and +k magnons were interchanged.

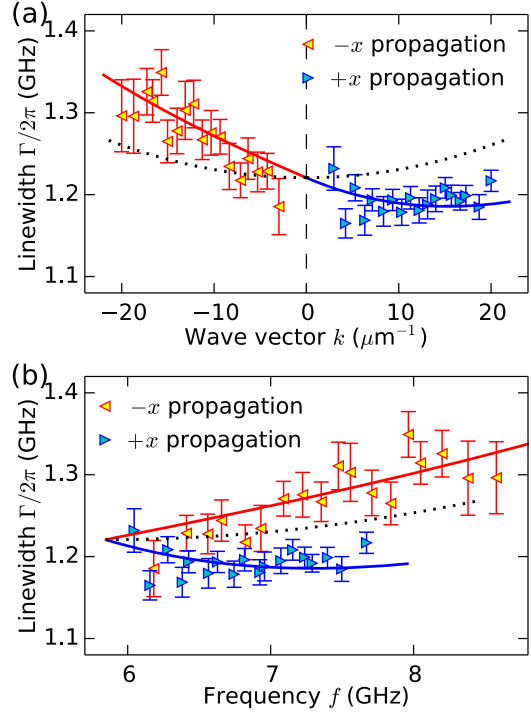

FIG. 4. Linewidth (FWHM) of spin wave in the Pt/Co/Ni film as a function of its (a) wave vector and (b) frequency for $H_0 = 54 \text{ mT}/\mu_0$ applied along the $-z$ direction. Experimental linewidths are denoted by triangles, with the error bars representing one standard deviation of the spectral fitting. The solid and dotted curves represent the calculated linewidths in the presence and absence of DMI, respectively.

To quantitatively explain the observed behavior of the magnon linewidth, the linearized LLG equation was solved with the Gilbert damping term present. Neglecting high-order terms $O(\alpha^2)$, the imaginary part of the angular frequency is

$$\text{Im}[\omega] = \alpha \gamma \mu_0 \left( H_0 + Jk^2 - H_U/2 + M_S/2 \right) \left[ 1 + \frac{\omega_{\text{DM}}(k)}{\omega_0(k)} \right], \quad (8)$$

with the real part $\text{Re}[\omega]$ given by Eq. (6). The Brillouin intensity is [42]

$$I(\omega') \propto \left| m(\omega', k) \right|^2 \propto \frac{1}{(\omega' - \text{Re}[\omega])^2 + \text{Im}[\omega]^2}, \quad (9)$$

where $m(\omega',k)$ is the Fourier transform of $m(x,t)$, and thus, the linewidth of the spin-wave peak is $\Gamma = 2\,\text{Im}[\omega]$. Eq. (8) reveals that, in the presence of DMI, the magnon linewidth is modified by the term $\omega_{DM}/\omega_0$, which is antisymmetric in the wave vector and thus accounts for the observed asymmetric linewidths. The calculated linewidths, as well as those in the absence of DMI, are plotted in Fig. 4. Good agreement with Brillouin measurements was obtained by setting $\alpha = 0.055$, which is consistent with literature values for Pt/Co thin films [43]. Interestingly, due to the presence of the interfacial DMI, the calculated linewidth has an unusual non-monotonic dependence on frequency for SWs propagating in the +x direction [see Fig. 4(b)].

In summary, direct experimental evidence of the existence of interfacial DMI in a Pt(4nm)/Co(1.6nm)/Ni(1.6nm) film was obtained by studying its spin-wave dynamics using Brillouin light scattering. The observed asymmetric magnon dispersion curves are attributed to an interfacial DMI-induced linear term in the magnon wave vector, which lifts the degeneracy of the magnon spectra. Such an asymmetry can be exploited in the design of nonreciprocal devices, such as spin-wave isolators, for magnonics and microwave technology applications. The DMI constant of the film was evaluated to be approximately 0.44 mJ/m$^2$. We have also experimentally demonstrated that the linewidths of counter-propagating magnons are different, with the difference being more pronounced for larger wave vectors, and have ascribed the difference to a DMI-induced term that is antisymmetric in the wave vector. Interestingly, our analytical calculations show that due to the existence of the DMI, the magnon linewidth is no longer a monotonic function of frequency. Our findings would be useful for understanding the influence of DMI on the domain-wall and spin-wave dynamics at Pt/ferromagnet interfaces.

This project was funded by the Ministry of Education, Singapore under Academic Research Fund Grant No. R144-000-340-112 and the National Research Foundation, Prime